\documentclass[sigconf]{acmart}

\usepackage[T1]{fontenc}
\usepackage{CJK}

\usepackage{kotex}
\usepackage{subfigure}
\usepackage{multirow}

\AtBeginDocument{%
  \providecommand\BibTeX{{%
    \normalfont B\kern-0.5em{\scshape i\kern-0.25em b}\kern-0.8em\TeX}}}

\copyrightyear{2025}
\acmYear{2025}
\setcopyright{rightsretained}
\acmConference[CHI EA '25]{Extended Abstracts of the CHI Conference on
Human Factors in Computing Systems}{April 26-May 1, 2025}{Yokohama, Japan}
\acmBooktitle{Extended Abstracts of the CHI Conference on Human Factors in
Computing Systems (CHI EA '25), April 26-May 1, 2025, Yokohama,
Japan}\acmDOI{10.1145/3706599.3720293}
\acmISBN{979-8-4007-1395-8/2025/04}

\begin{document}

\begin{CJK}{UTF8}{mj}

\title[Concert Interaction Translation]{Concert Interaction Translation: Augmenting VR Live Concert Experience using Chat-Driven Artificial Collective Reactions}

\author{Sebin Lee}
\email{leesebin@soongsil.ac.kr}
\orcid{0000-0002-8927-0319}
\affiliation{%
  \institution{Department of Culture Contents,\\Soongsil University}
  \city{Seoul}
  \country{Republic of Korea}
}

\author{Yeonho Cho}
\email{ckc2051@soongsil.ac.kr}
\orcid{0009-0001-2671-8593}
\affiliation{%
  \institution{Department of Culture Contents,\\Soongsil University}
  \city{Seoul}
  \country{Republic of Korea}
}

\author{Jungjin Lee}
\email{jungjinlee@ssu.ac.kr}
\orcid{0000-0003-3471-4848}
\affiliation{%
  \institution{The Global School of Media,\\Soongsil University}
  \city{Seoul}
  \country{Republic of Korea}
}
\authornote{Corresponding author}

\renewcommand{\shortauthors}{Sebin Lee, Yeonho Cho, and Jungjin Lee}

\begin{abstract}
    Computer-mediated concerts can be enjoyed on various devices, from desktop and mobile to VR devices, often supporting multiple devices simultaneously.
    However, due to the limited accessibility of VR devices, relatively small audience members tend to congregate in VR venues, resulting in diminished unique social experiences.
    To address this gap and enrich VR concert experiences, we present a novel approach that leverages non-VR user interaction data, specifically chat from audiences watching the same content on a live-streaming platform.
    Based on an analysis of audience reactions in offline concerts, we designed and prototyped a concert interaction translation system that extracts the level of engagement and emotions from chats and translates them to collective movements, cheers, and singalongs of virtual audience avatars in a VR venue.
    Our user study (n=48) demonstrates that our system, which combines both movement and audio reactions, significantly enhances the sense of immersion and co-presence than the previous method.
\end{abstract}

\begin{CCSXML}
<ccs2012>
   <concept>
       <concept_id>10003120.10003121.10003128</concept_id>
       <concept_desc>Human-centered computing~Interaction techniques</concept_desc>
       <concept_significance>500</concept_significance>
       </concept>
    <concept>
        <concept_id>10003120.10003121.10003122.10003334</concept_id>
        <concept_desc>Human-centered computing~User studies</concept_desc>
        <concept_significance>500</concept_significance>
    </concept>
    <concept>
        <concept_id>10003120.10003121.10003124.10010866</concept_id>
        <concept_desc>Human-centered computing~Virtual reality</concept_desc>
        <concept_significance>500</concept_significance>
    </concept>
 </ccs2012>
\end{CCSXML}

\ccsdesc[500]{Human-centered computing~Interaction techniques}
\ccsdesc[500]{Human-centered computing~Virtual reality}
\ccsdesc[500]{Human-centered computing~User studies}

\keywords{VR concerts, live-streaming concerts, audience reactions, human-computer interaction, user studies}

\begin{teaserfigure}
  \includegraphics[width=\textwidth]{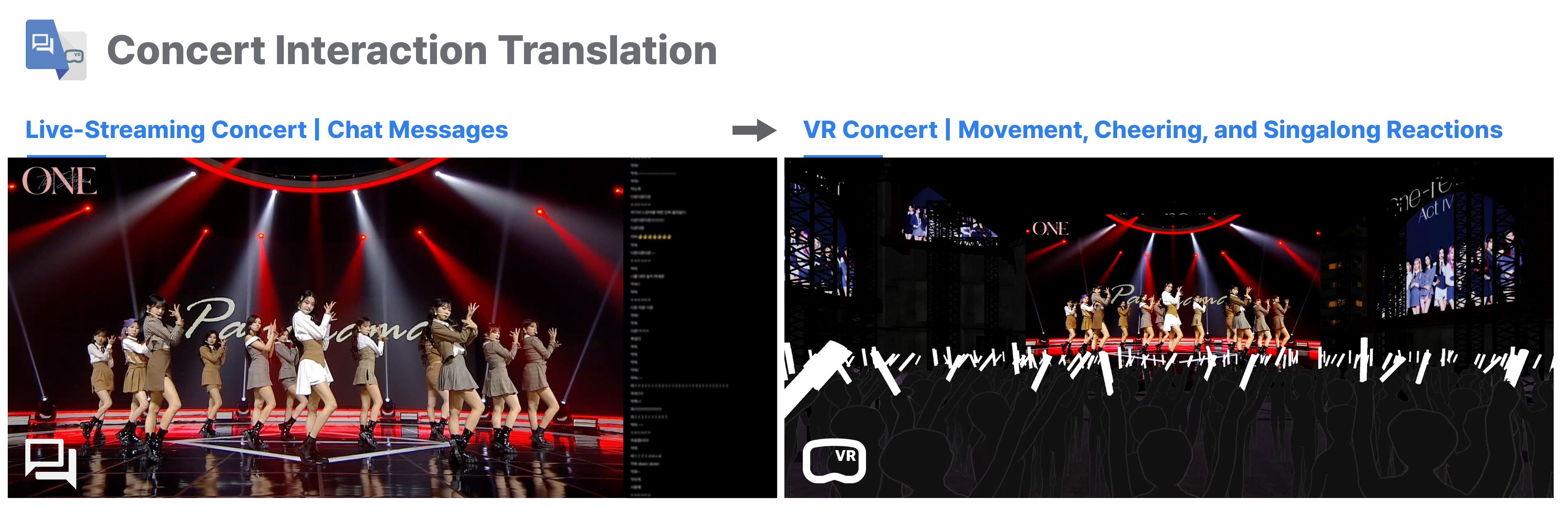}
  \caption{Concept of Concert Interaction Translation (CIT) system: The CIT system analyzes chat interactions produced by live-streaming audience members in real time and translates them into the collective reactions of dummy audience avatars in a VR concert venue to enhance the audience experience in VR.}
  \Description{Conceptual Visualization of the Concert Interaction Translation System. Two side-by-side images compare a live-streamed concert performance (left) with chat messages on-screen, and a virtual reality concert setting (right) featuring audience avatars. The figure highlights how real-time chat interactions are translated into avatar reactions—movement, cheering, and singalong—to enhance immersion in the VR venue.}
  \label{fig:teaser}
\end{teaserfigure}

\maketitle

\section{Introduction}
Live online concerts have emerged as a crucial platform for fandom businesses, particularly following the COVID-19 pandemic.
They have also become an essential medium for newer entertainers, such as virtual YouTubers, to showcase their talents \cite{JuTaime, Concert-VTuber-1, Concert-VTuber-2}.
These concerts are delivered through various methods, including livestreaming on platforms like YouTube and Twitch, where audiences engage via chat \cite{ConcertExp-LiveStreamChat-2, ConcertExp-LiveStream-2}, as well as through VR venues like VRChat, which offer immersive experiences using head-mounted displays (HMDs) \cite{MMConcert, JuTaime}.
Recently, some performers have adopted multiple delivery methods simultaneously to cater to diverse audience preferences \cite{EnhancingExp-MultiPlatform-1, EnhancingExp-MultiPlatform-2, JuTaime}.

However, the delivery methods reveal a disparity in audience participation.
Livestreams are widely accessible and tend to attract large audiences, while VR concerts, although highly immersive, appeal to a niche group due to the requirement of HMDs \cite{JuTaime}.
Furthermore, VR platforms often limit the number of users per session to ensure technical stability.
These limitations result in smaller audience sizes in VR, which diminishes the potential to cultivate the social experiences crucial for enhancing audience engagement in the entertainment domain \cite{ConcertExp-Offline-7, Mulder-2023-Constructing}.

Researchers have explored various computational methods to address the lack of audiences in VR musical performances.
One common approach involves augmenting visual elements, such as adding non-player character (NPC) audiences or triggering visual effects based on biosignals that estimate emotional states \cite{VRConcertReview, EnhancingExp-VFXBio-2, EnhancingExp-VFXBio-3, EnhancingExp-VFXBio-4, EnhancingExp-VFXBio-5, EnhancingExp-VFXBio-6}.
While these methods provide emotionally accurate reflections, they require additional sensing hardware, limiting their practicality in real-world applications.
Other approaches have focused on leveraging contextual information within the concert itself.
For example, Yakura and Goto proposed synthesizing cheering motions of dummy avatars based on the acoustic features of concert music \cite{Baseline}.
However, the emotions conveyed by music do not always align with the audience’s actual feelings \cite{Kawakami-2012-Relations, Evans-2008-relationships}.
This misalignment can result in artificial audience reactions that differ from the true emotional states of the participants.
Furthermore, previous studies have primarily focused on leveraging visual elements, overlooking auditory responses like cheers and singalongs, which are integral to offline concert experiences.

In this paper, inspired by the observation that live online concerts often utilize multiple delivery methods simultaneously \cite{EnhancingExp-MultiPlatform-1, EnhancingExp-MultiPlatform-2, JuTaime}, we introduce the Concert Interaction Translation (CIT) system that uses audience reactions from live-streaming concerts to enhance VR live concert experiences (Figure \ref{fig:teaser}).
The system translates chat interactions into the collective reactions of dummy audience avatars in VR, replicating both verbal and non-verbal audience responses commonly seen in offline concerts.
To achieve this, we first analyzed 68 concert videos to identify representative movement and sound reactions based on the emotional arousal and valence of the audience.
Subsequently, we designed the CIT system that estimates engagement and emotional level from live-streaming concert chats in real time and synthesizes movement, cheers, and singalongs in VR.
A user study using a proof-of-concept prototype of CIT demonstrated that the CIT significantly enhances immersion and co-presence compared to the previous method that relied solely on acoustic features of concert music \cite{Baseline}.
\begin{table*}[hbt!]
\caption{Description of classified movement and sound reactions according to the arousal and valence level}
\Description{Description of Movement and Sound Reactions by Arousal and Valence. This table categorizes audience reactions into three arousal levels (high, neutral, low) and three valence levels (negative, neutral, positive). It outlines specific arm movements—such as shaking arms back and forth or side to side—and vocal expressions—like cheering or disappointed sounds—based on each combination of arousal and valence, with references to relevant figures and external timestamps.}
\label{tab:audience-reaction}

\begin{center}
\begin{tabular}{|p{0.08\textwidth}|p{0.3\textwidth}|p{0.25\textwidth}|p{0.25\textwidth}|}
\hline

% Row 1
\parbox{0.08\textwidth}{
    \centering
    \vspace{5pt}
    High\\Arousal
    \vspace{5pt}
}&

% Low Valence Content
\multirow{3}{0.3\textwidth}{%
  \parbox{0.3\textwidth}{%
    \centering
    \textbf{Movement:}\\
    Disappointing Motion (Figure~\ref{fig:motion-disappointing})\\[4pt]
    \textbf{Sound:}\\
    Disappointing Sound: \textit{Ahhhh\ldots}\\
    (cf.~\cite{F57} (0:42--0:50))
  }
}&

% High Arousal & Neutral/High Valence
\multicolumn{2}{p{0.5\textwidth}|}{%
  \parbox{0.5\textwidth}{%
    \centering
    \vspace{5pt}
    
    \textbf{Movement:}\\
    \textit{In fast tempo:} Shaking Arms: Back and Forth (Figure~\ref{fig:motion-bnf})\\
    \textit{In low tempo:} Shaking Arms: Side by Side (Figure~\ref{fig:motion-sbs})\\
    \textit{When talking toward audience:} Hand clap (Figure~\ref{fig:motion-handclap})\\[4pt]
    \textbf{Sound:}\\
    Shouting: \textit{``Waaaaa,'' ``Woooow,'' or ``Hoooou''}\\
    (cf.~\cite{F2, F36} (0:00--0:10))
    \vspace{5pt}
  }
}
\\
\cline{1-1}\cline{3-4}

% Row 2
\parbox{0.08\textwidth}{
    \centering
    \vspace{5pt}
    Neutral\\Arousal
    \vspace{5pt}
}&&

% Neutral Arousal - Neutral/High Valence
\multicolumn{2}{p{0.5\textwidth}|}{%
  \parbox{0.5\textwidth}{%
    \centering
    \textbf{Movement:} Hand clap (Figure~\ref{fig:motion-handclap})\\
    \textbf{Sound:} Hand clap (cf.~\cite{F55} (0:30--0:40))
  }
}
\\

% Make a line
\cline{1-1}\cline{3-4}

% Row 3
\parbox{0.08\textwidth}{
    \centering
    \vspace{5pt}
    Low\\Arousal
    \vspace{5pt}
}
&
& \multicolumn{2}{c|}{Not Assigned}
\\
\hline

% Bottom row of valence labels
& \parbox{0.3\textwidth}{
    \centering 
    \vspace{5pt}
    Low Valence\\(Negative Emotion)
    \vspace{5pt}    
}
& \parbox{0.25\textwidth}{
    \centering 
    \vspace{5pt}
    Neutral Valence
    \vspace{5pt}    
}
& \parbox{0.25\textwidth}{
    \centering 
    \vspace{5pt}
    High Valence\\(Positive Emotion)
    \vspace{5pt}    
}
\\
\hline

\end{tabular}
\end{center}
\end{table*}

\section{Analysis of Audience Reactions in Offline Concerts}
\label{sec:AudienceReactions}
We first analyzed offline concert footage to understand audience reactions in the wild. 
We collected footage from YouTube using keywords such as ``Live Concert,'' ``Concert,'' and ``Concert Fancam.''
Additionally, we included a complete concert video from a Blu-ray disc to examine reactions during singing, performances, and conversations.
We gathered 68 videos spanning 12 hours and 25 minutes and extracted 3 hours and 9 minutes of audience-focused clips.
These clips were qualitatively analyzed using Russell's circumplex model, which categorizes emotions along two dimensions: arousal (intensity) and valence (positivity or negativity) \cite{RussellEmotionTheory}, a framework commonly used in prior music studies to analyze audience emotions \cite{Egermann-2013-Probabilistic, Theorell-2019-Emotional}.

\subsection{Movement Reactions}
\label{subsec:cheering-motion}

\begin{figure}
    \centering
    \subfigure[Shaking Arm: Back\&Forth]{
        \includegraphics[width=.45\linewidth]{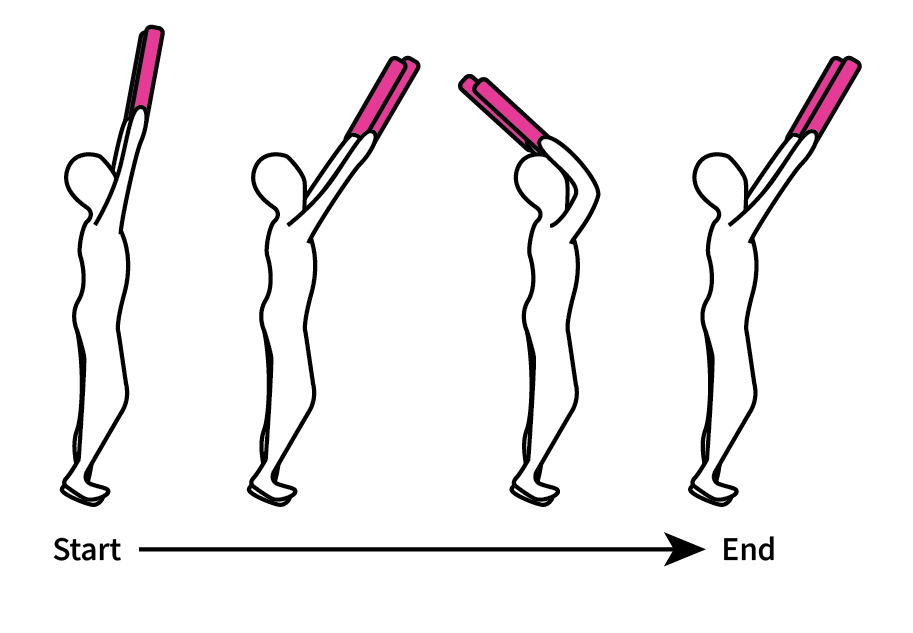}
        \label{fig:motion-bnf}
    }
    \subfigure[Shaking Arm: Side by Side]{
        \includegraphics[width=.45\linewidth]{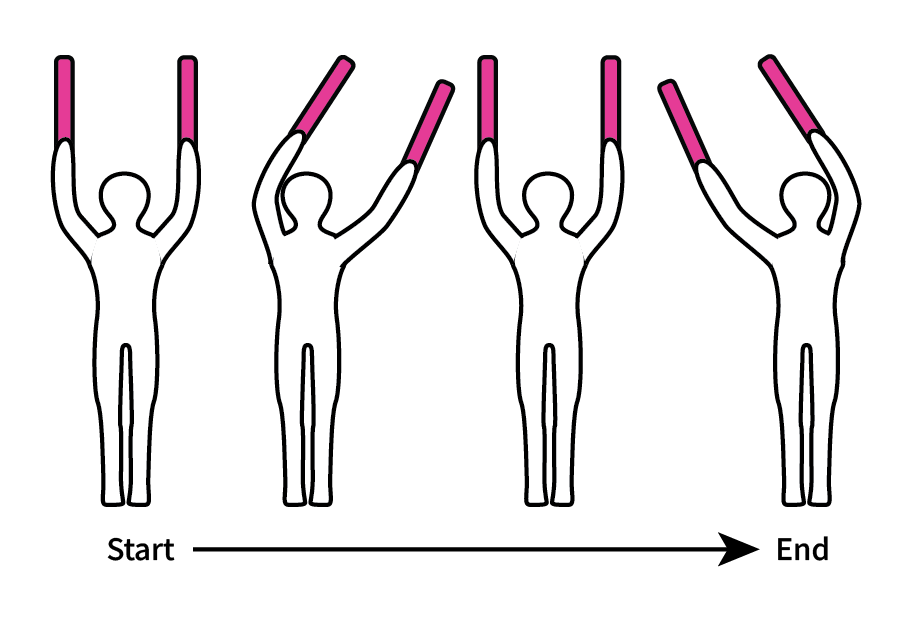}
        \label{fig:motion-sbs}
    }
    \subfigure[Hand clap]{
        \includegraphics[width=.45\linewidth]{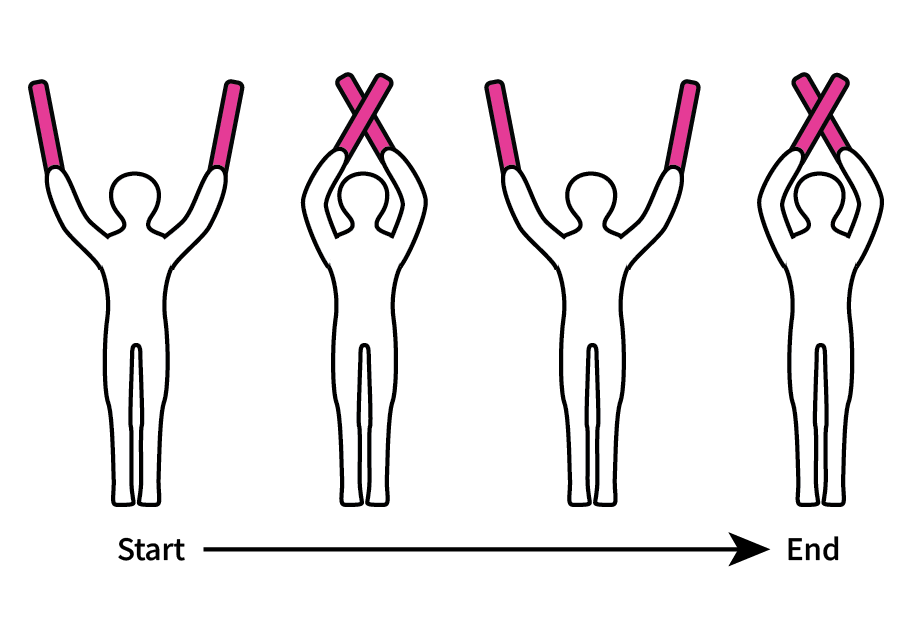}
        \label{fig:motion-handclap}    
    }
    \subfigure[Disappointing Motion]{
        \includegraphics[width=.45\linewidth]{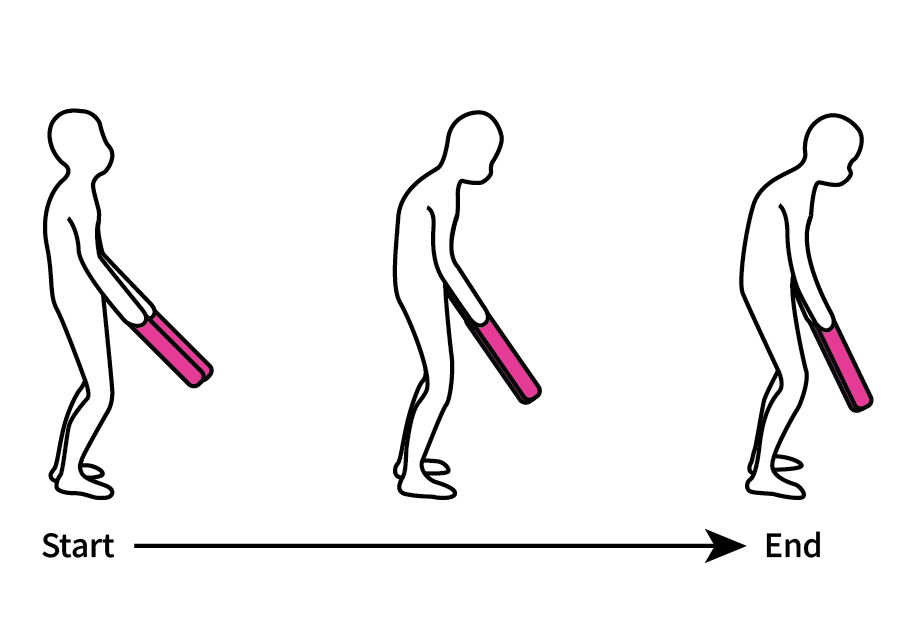}
        \label{fig:motion-disappointing}
    }
    \caption{Description of observed movement reactions}
    \Description{Illustration of Four Distinct Audience Movement Reactions. This figure depicts four silhouettes, each demonstrating a different audience movement. From left to right: (a) arms shaking back and forth, (b) arms shaking side by side, (c) a hand clap, and (d) a disappointed posture. Each sub-figure shows start and end positions, highlighting how audience members vary their motions in response to concert events.}
    \label{fig:audience-movement}
\end{figure}

Our analysis of audience movement reactions aligns with previous studies, emphasizing beat-synchronized motions as a form of artist support \cite{Baseline, AudienceReaction-1}.
Most audience members exhibited uniform motions, such as arm-waving in unison, with some variations, like shaking arms in different directions.
Emotional responses were predominantly positive and intense, with negative emotions like anger or disappointment rarely observed.

During moments of high arousal and positive valence, two main arm movements were identified: swaying arms back and forth (Figure \ref{fig:motion-bnf}, cf. \cite{F52} (0:20-end), \cite{F3} (2:01-2:14)) or side to side (Figure \ref{fig:motion-sbs}, cf. \cite{F15} (3:33-3:40), \cite{F41} (3:03-3:14)), depending on the music tempo.
Fast songs typically elicited back-and-forth movements, while slower songs prompted side-to-side swaying (Table \ref{tab:audience-reaction}).
Clapping, indicating moderate arousal, was common during performer-audience interactions (Figure \ref{fig:motion-handclap}, cf. \cite{F55} (0:30-0:40)).
Negative emotions, such as disappointment, were difficult to observe due to poor lighting. We designed a `disappointed' motion to represent such emotions based on a researcher's concert experience (Figure \ref{fig:motion-disappointing}).
Calm, low-intensity positive emotions did not result in notable physical reactions.

\subsection{Sound Reactions}
\label{subsec:cheering-sound}
Alongside movements, audiences actively participated in vocal expressions of support to bolster the performers.
The sound reactions can be categorized into cheering and singalongs.

\subsubsection{Cheering Sounds}
Cheering, a non-lyrical vocal response, typically occurs during popular songs or impressive musical solos, consistent with prior research \cite{AudienceReaction-2}.
The nature of cheers varies based on audience emotions (Table \ref{tab:audience-reaction}).
High-arousal positivity elicits loud, high-pitched cheers like "Waaa" or "Wow" (cf. \cite{F2, F36} (0:00–0:10)), while neutral arousal prompts applause (cf. \cite{F55} (0:30–0:40)).
Disappointment or negative emotions result in sighs or groans, such as "Ahhhh" (cf. \cite{F57} (0:42–0:50)).
Similar to movements, vocal reactions are absent during low-intensity positive emotions.

\subsubsection{Singalong Sounds}
\label{subsubsec:singalong}
Singalongs, including fan chants, involve audiences singing lyrics or reciting chants at specific moments (henceforth, we will refer to these terms collectively as `singalong').
Singalongs typically occur at highlight parts of songs, and fan chants follow predefined guidelines provided by artists or fandoms, performed at designated moments \cite{Fanchant-1, Fanchant-2}.
These responses are organized and occur regardless of the emotional atmosphere (cf. \cite{F29} (2:03–2:25), \cite{F14} (2:57–3:10), \cite{F46} (16:32–16:50)).

\section{Concert Interaction Translation}
\begin{figure*}
    \centering
    \includegraphics[width=\linewidth]{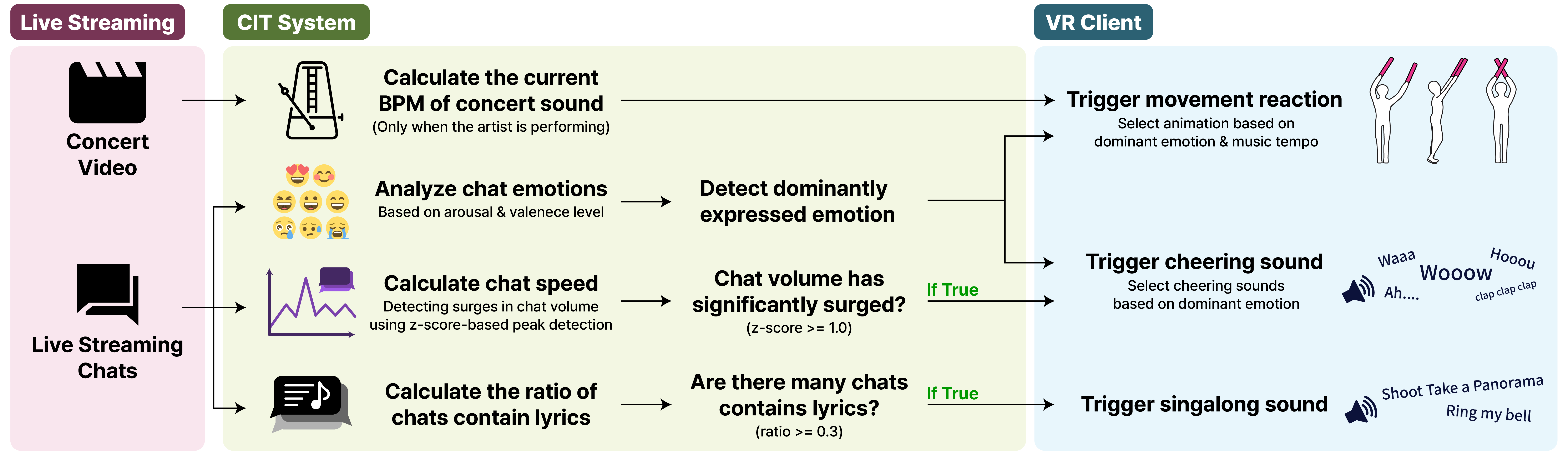}
    \caption{Overview of the proposed system, Concert Interaction Translation (CIT)}
    \label{fig:system-overview}
    \Description{Overview of the Proposed Concert Interaction Translation (CIT) System. This figure shows a pipeline starting with concert video and live-streaming chat inputs on the left, processed by the CIT system in the middle to calculate music tempo and analyze chat volume and emotion. On the right, a virtual reality client responds by triggering movement, cheering, or singalong based on these analyses.}
\end{figure*}
Drawing from the audience reactions identified from offline concert videos, we developed the following design goals in leveraging live-streaming chat data to augment VR concert venues with artificial audience reactions:

\begin{itemize}
    \item Initiate suitable movement reactions that reflect the emotional states of live-streaming concert viewers.
    \item Align audience reactions with precise timing to ensure a cohesive and authentic concert experience.
\end{itemize}

As per the design goals, we designed the Concert Interaction Translation (CIT) system.
Figure \ref{fig:system-overview} presents an overview of the CIT method, which provides VR concert audience with movement reactions, cheering sounds, and singalong reactions based on the chat interaction in a live-streaming concert.
Our method involves analyzing the collective engagements and emotions expressed by audience chats during live-streaming concerts and mapping the analyzed results to three different types of reactions.

\subsection{Triggering Movement Reactions}
The CIT server collects chat messages from the live-streamed concert, estimates their emotions, and categorizes the predominant emotion as high, neutral, or low arousal and valence levels, sending the emotion level to the client.
On the client side, the system determines the song's tempo by evaluating its beats per minute (BPM) to determine the song's tempo.
If the BPM is above 100 BPM, the tempo is considered fast, and if it is below, it is considered slow, based on empirical data for the prototype.
The client triggers movement reactions based on the tempo and emotion described in Table \ref{tab:audience-reaction} and adjusts animation speeds to synchronize dummy avatars' movements with the song rhythm.
During artist-audience dialogue, beat synchronization is disabled to avoid mismatched animation speeds.
To distinguish the performance and conversation, we predefined timestamps that indicate performance and conversation periods.

Movement reaction animations were created using motion capture (Figure \ref{fig:audience-movement}) and refined by manually editing keyframes to remove inconsistencies.
To introduce variation, we modified animations, such as changing arm movements or raising one arm instead of both.
The CIT system assigns these variations with a 20\% chance to dummy audience avatars in the virtual concert venue.

\subsection{Triggering Cheering Sounds}
The system uses chat speed and emotions to trigger cheering sounds by continuously monitoring chat frequency second by second.
A significant surge in chat volume prompts the system to analyze dominant emotions and trigger corresponding cheering sounds.
We used a moving average z-score-based peak detection algorithm \cite{PeakDetection}, applied in a previous live-streaming study \cite{PeakDetection-1} to detect these surges.
We empirically set a 5-second moving average window for the CIT prototype and triggered the sounds when the z-score exceeded 1.0.

The cheering sounds used in the system were carefully chosen by a researcher with more than ten years of experience attending concerts.
The researcher selected appropriate open-source cheering sound assets tailored to the concert setting and also extracted cheering sounds directly from concert footage to ensure authenticity and alignment with the dynamic atmosphere of a live concert.

\subsection{Triggering Singalong Sounds}
\label{subsec:triggering-singalong-fanchant}
As explained in Section \ref{subsubsec:singalong}, singalongs are timed events that occur at predetermined moments within a song.
For the prototype, we predefined timestamps indicating potential singalong points.
When the concert reaches one of these designated intervals, the system scans for chats containing lyrics using regular expressions.
If the ratio of lyric-containing chats to total chats at the time surpasses a predetermined threshold, which was empirically set at 30\%, a singalong sound is initiated.
To implement the prototype, we had four undergraduate students record singalong sounds under the supervision of a researcher with ten years of concert experience.
We refined these recordings using Adobe Audition to produce sound effects that closely resemble the chorus of a crowd engaged in a singalong.

\subsection{Implementation of the Prototype System}
\begin{figure*}[hbt!]
    \centering
    \includegraphics[width=\linewidth]{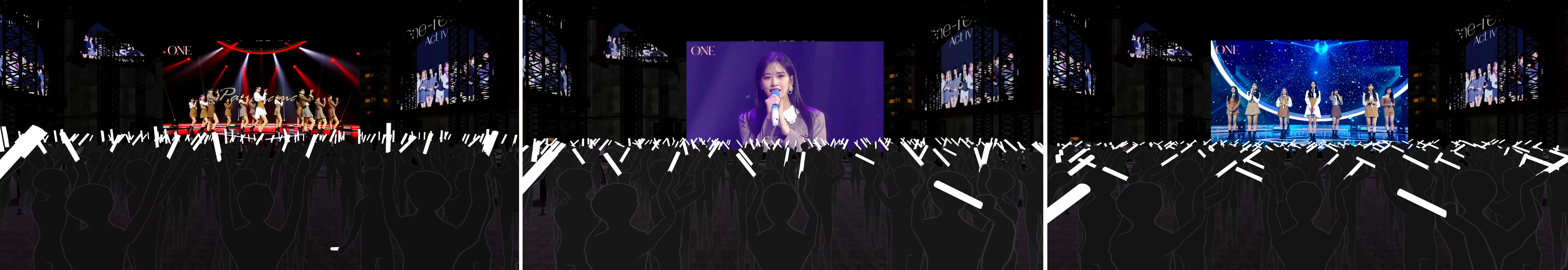}
    \caption{Images captured from the CIT prototype (left: shaking arms back and forth to a fast-tempo song, center: clapping hands during a talk session, right: clapping hands to a slow-tempo song)}
    \label{fig:cit-example}
    \Description{Images from the CIT Prototype Showing a Virtual Concert Venue. Three screenshots display a front screen with a recorded concert and rows of audience avatars. On the left, avatars shake their arms back and forth during a fast-tempo song. In the center, avatars clap hands during a talk session. On the right, avatars clap hands to a slow-tempo song.}
\end{figure*}

We implemented the CIT system prototype for the user study.
A recorded video and chat data from the live-streamed "ONE, THE STORY" concert \cite{OneTheStory}, performed on March 11, 2021, were used to create a controlled experimental setting.
Given that the majority of participants were anticipated to be Korean, we selected the ``ONE, THE STORY'' concert \cite{OneTheStory} performed by the famous Korean idol group IZ*ONE.

We developed a server using Python's FastAPI framework to process live-streaming chat data and transmit the analyzed results.
The server reads chat data from a CSV file at one-second intervals, processes the data following the workflow in Figure \ref{fig:system-overview}, and transmits the processed data to the client via WebSocket.
We utilized the KOTE model for emotion classification of Korean chat data, demonstrating high performance in classifying emotions in Korean text \cite{KOTE}.
The KOTE model categorizes text into 44 distinct emotions but does not directly provide arousal and valence values \cite{KOTE}.
Therefore, we classified the valence of the 44 emotions into high (positive), neutral, and low (negative) based on the valence levels described in the KOTE paper \cite{KOTE}.
Similarly, the arousal levels of these emotions were classified into high, neutral, and low using Lee's study on Korean emotion classification \cite{KoreanEmotion} as a reference.

We built the virtual concert venue using Unity 2022.3.10f.
As shown in Figure \ref{fig:cit-example}, the venue features a front screen displaying the recorded concert video and 200 dummy audience avatars with light sticks positioned between the screen and the user at the rear of the crowd, similar to prior studies \cite{MMConcert, Baseline}.
To estimate the song's BPM, we integrated the RhythmTool asset, widely used for music analysis in Unity projects \cite{RhythmTool}.

For the user study, we deployed the server on a high-performance workstation with an Intel Xeon Gold 5218R CPU, 256 GB memory, and dual NVIDIA RTX A6000 GPUs to ensure smooth processing.
The VR client, installed on a Dell G15 5520 laptop with an RTX 3070Ti GPU, was connected to the server via a stable local area network to avoid latency issues.

\section{User Study}
\subsection{Study Design}
The objective of the user study was to compare our method with the previous method and to evaluate how different types of artificial reactions affect the audience experience in a VR live concert.
To achieve this, we developed four scenarios, each with different conditions:

\begin{itemize}
    \item S1: Yakura and Goto's approach (synthesized movement reaction based on acoustic features) \cite{Baseline}
    \item S2: Movement reaction
    \item S3: Movement + cheering sound reaction
    \item S4: Movement + cheering sound + singalong sound reaction
\end{itemize}

It should be noted that we reimplemented the Yakura and Goto’s method because the original codes or the executable files are not publicly available.
In each scenario, participants watched a 10-minute recorded live concert video within a virtual concert venue.
Dummy audience avatars reacted to the video based on the specific configuration of each scenario.
To encompass various situations commonly encountered in live concerts, the experimental video consisted of three parts: a fast-paced and energetic song (IZ*ONE - Panorama \cite{Panorama}), talking toward the audience, and a slow-paced and emotional song (IZ*ONE - With*One \cite{WithOne}).

\subsection{Procedure}
For the user study, we recruited 48 participants (27 females, 21 males) aged 18 to 32 years (M=22.1, SD=2.8). 
All Participants were invited to the laboratory for the experiment.
Prior to the experiment, the researchers explained the procedure and informed participants they could stop the experiment if they felt any discomfort.
Each participant experienced all scenarios in a counterbalanced order using a Meta Quest Pro HMD and Sennheiser HD 4.50 btnc headphones to minimize external noise.

After viewing each scenario, participants took off the HMD and completed a survey about their viewing experience.
The survey measures presence, co-presence, immersion, and sickness, in line with previous research on virtual music concerts \cite{VRConcertReview, JuTaime, MMConcert, Baseline}.
Sickness was assessed with the simulator sickness questionnaire (SSQ) \cite{SSQ}, presence with the igroup presence questionnaire (IPQ) \cite{IPQ}, and immersion with the FilmIEQ \cite{FilmIEQ}, a media-specific adaptation of the immersive experience questionnaire (IEQ) \cite{IEQ}.
Co-presence was measured using a co-presence questionnaire presented in Yakura and Goto's study \cite{Baseline}, which was adapted from Hwang and Lim's co-presence questionnaire \cite{Co-presence} specifically for the concert context.
Most metrics used a 7-point Likert scale, except sickness, which used a 4-point scale calculated per Kennedy et al. \cite{SSQ}.
The entire process took approximately 70 minutes.

\subsection{Result}
\begin{figure*}
    \centering
    \subfigure[Sense of sickness]{
        \includegraphics[width=.23\linewidth]{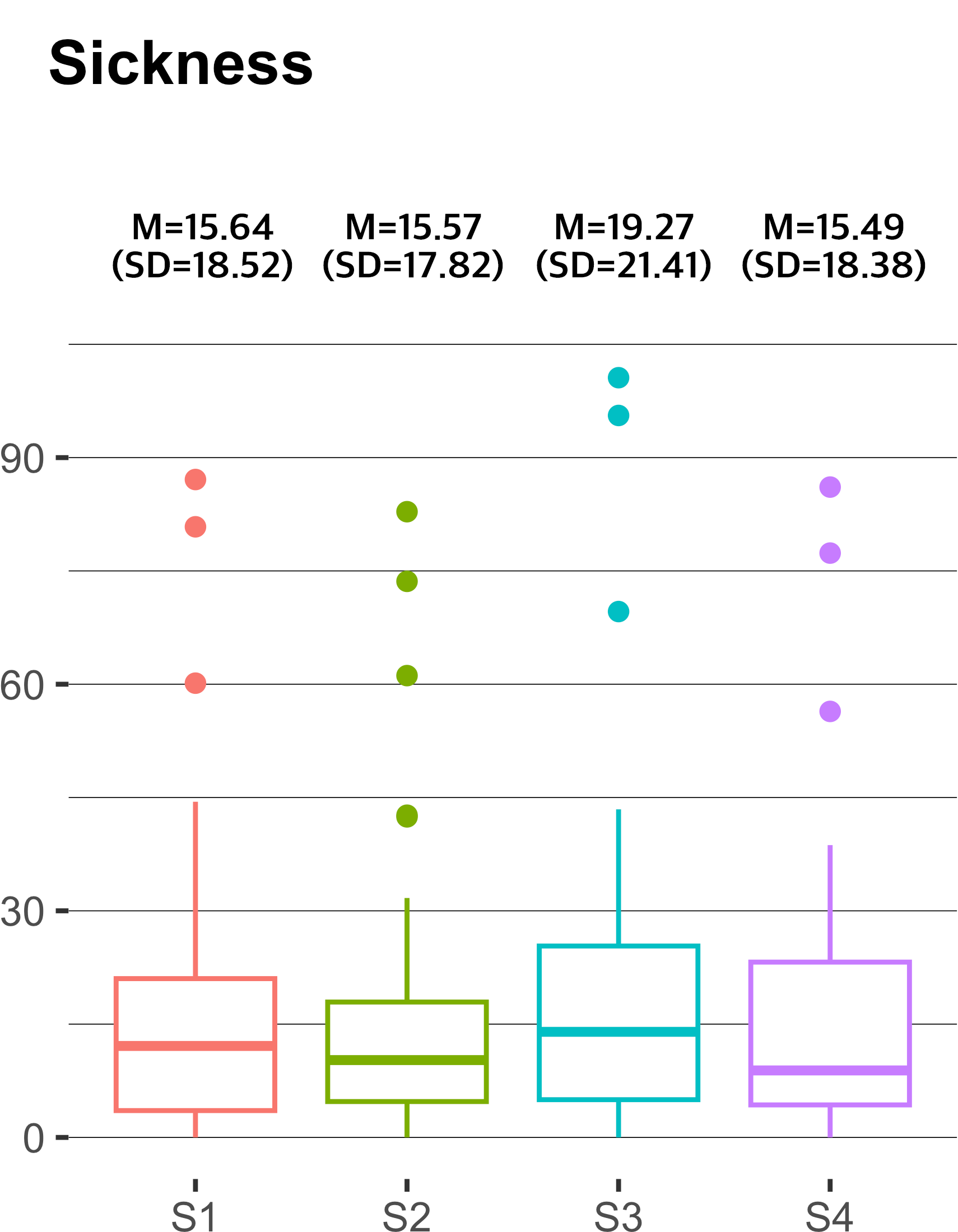}
        \label{fig:quantitative-sickness}
    }
    \subfigure[Sense of presence]{
        \includegraphics[width=.23\linewidth]{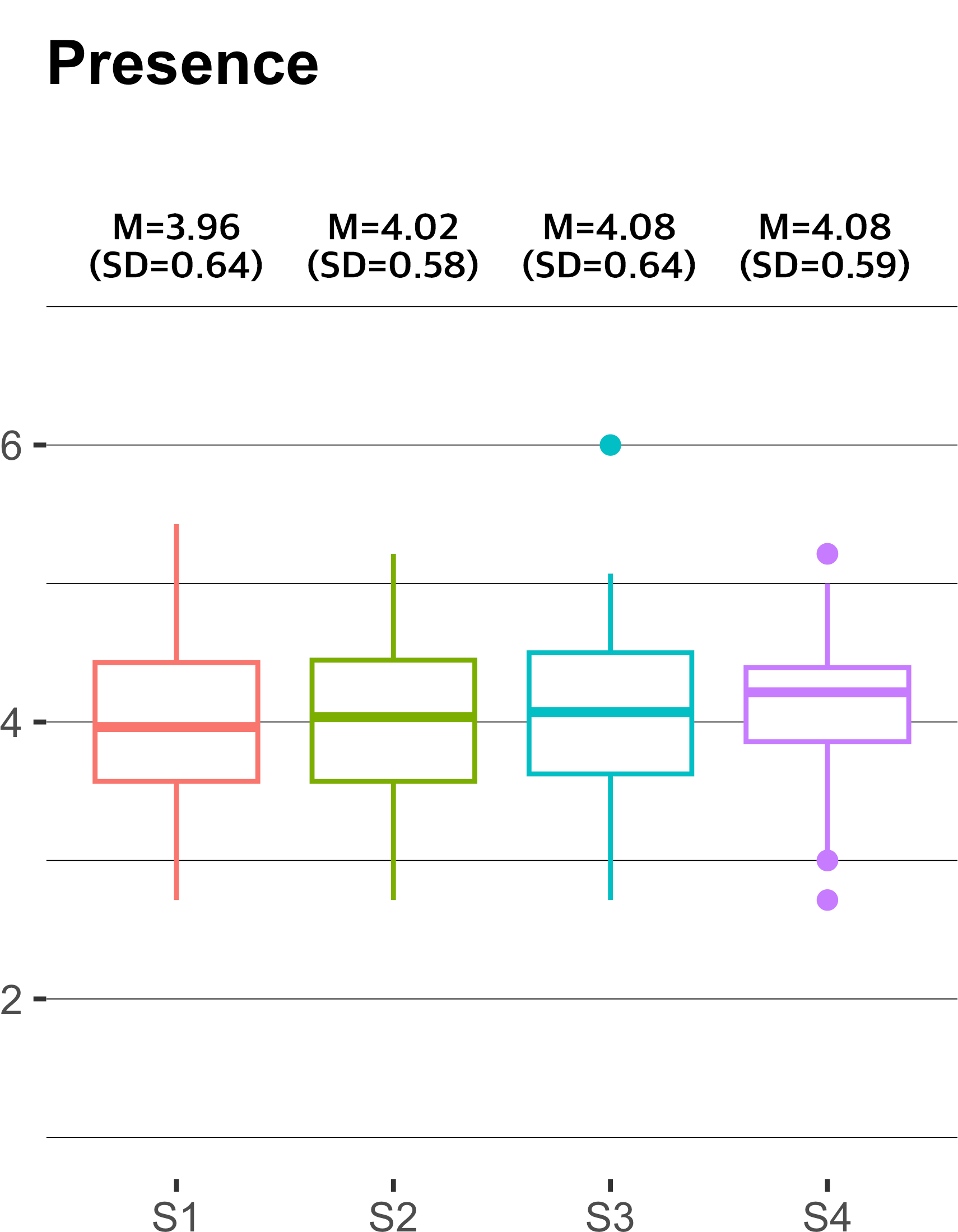}
        \label{fig:quantitative-presence}
    }
    \subfigure[Sense of immersion]{
        \includegraphics[width=.23\linewidth]{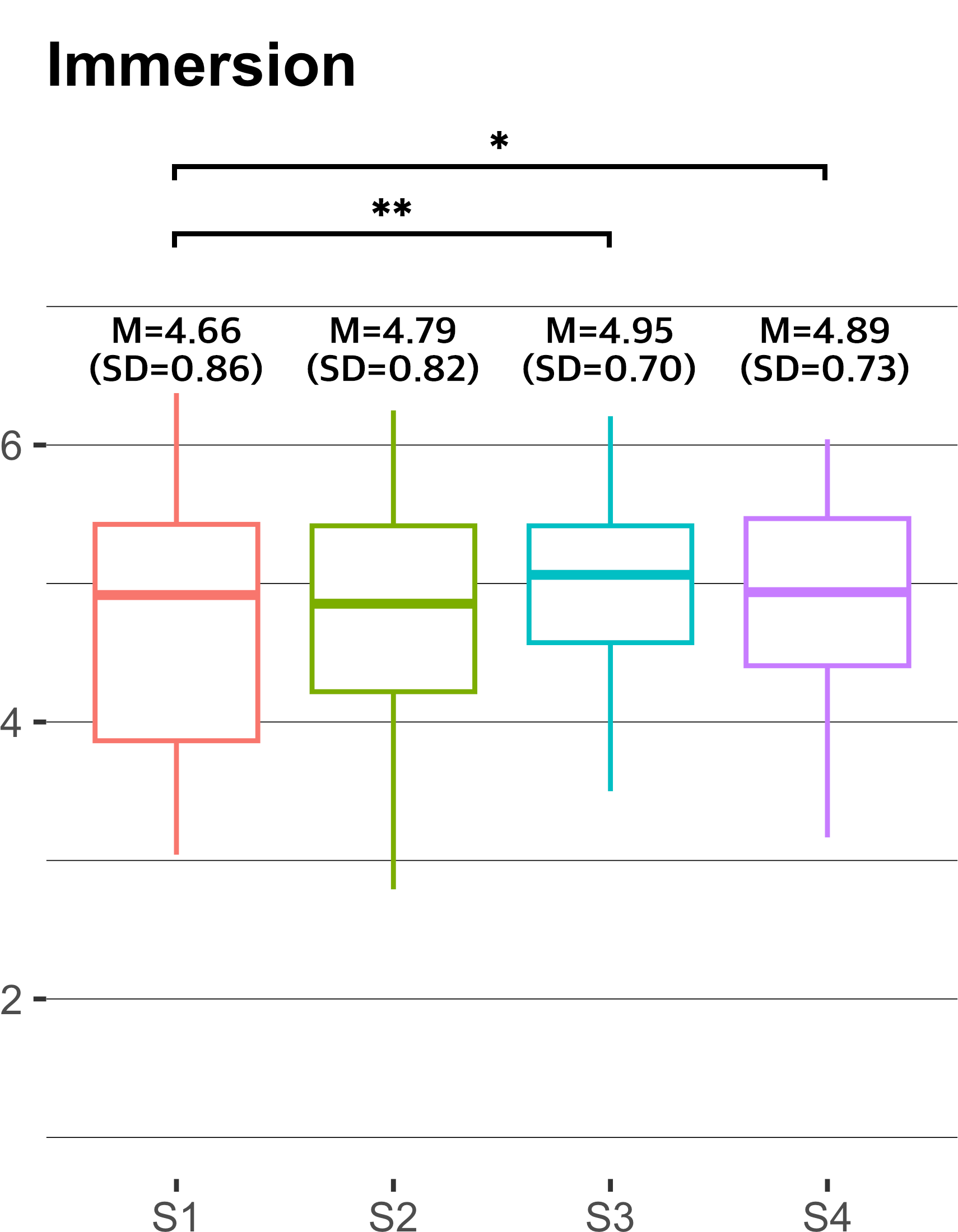}
        \label{fig:quantitative-immersion}
    }
    \subfigure[Sense of co-presence]{
        \includegraphics[width=.23\linewidth]{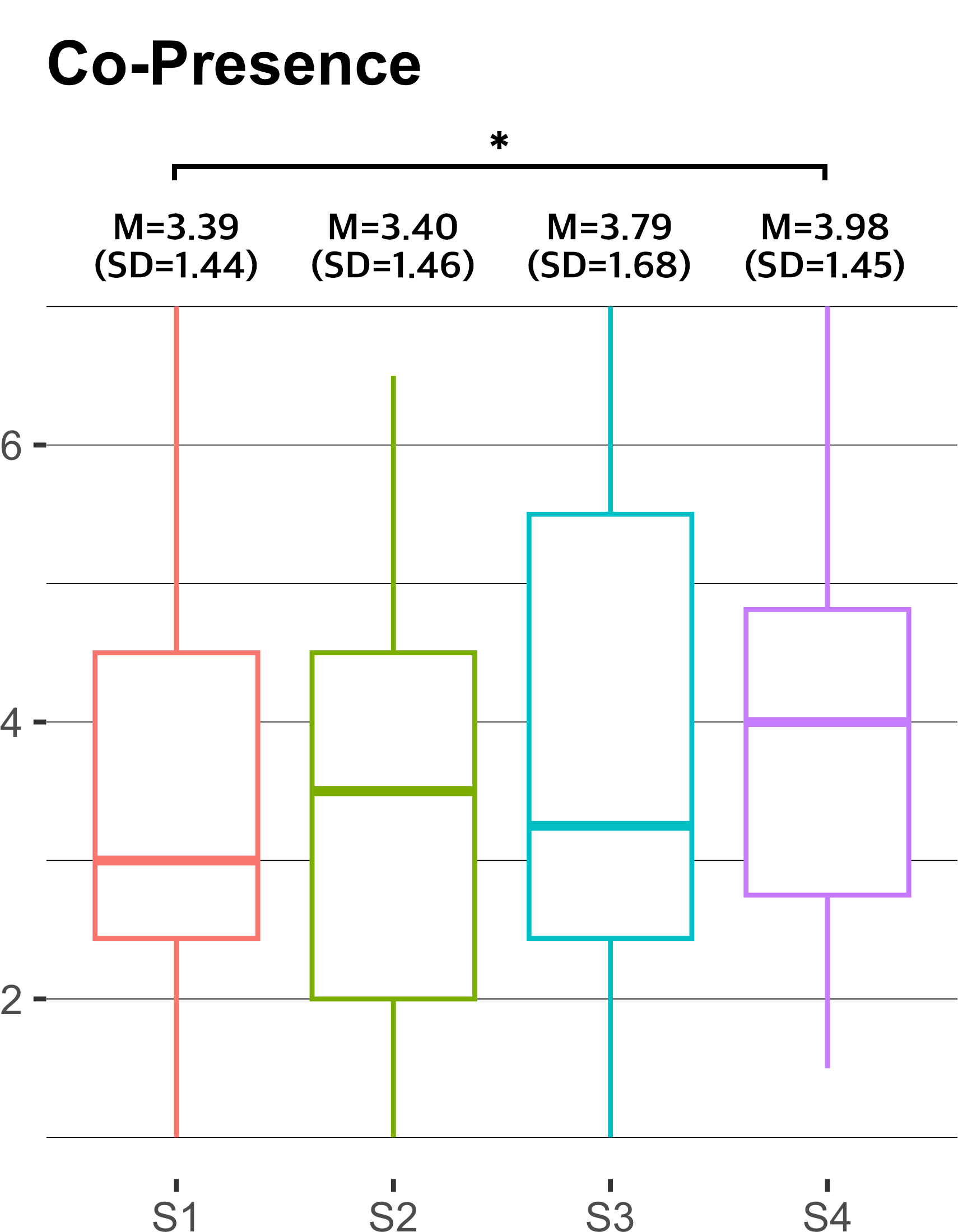}
        \label{fig:quantitative-copresence}
    }
    \caption{The results of the analysis on overall sickness, presence, immersion, and co-presence according to four scenarios (S1-S4). (signif. level: * $p<.05$, ** $p<.01$)}
    \Description{Box Plots of Sickness, Presence, Immersion, and Co-Presence for Four Scenarios (S1–S4). Four box plots compare participant ratings of sickness, presence, immersion, and co-presence across the scenarios. CIT system with sound reactions (S3 and S4) has indicated that higher immersion and co-presence than method proposed in previous study.}
    \label{fig:quantitative-graph}
\end{figure*}

To assess the viewing experience, we used a one-way repeated measures ANOVA, with the experimental scenarios as a within-subject factor.
When the data violated the assumption of sphericity as determined by Mauchly's test, we applied the Greenhouse-Geisser correction for conservative analysis.
When a significant difference was detected in the repeated measures ANOVA, post-hoc analysis was performed using the TukeyHSD test to identify specific group differences.

A significant difference was observed in sickness across scenarios (F(3, 141) = 3.268, $p < .05$).
However, post-hoc analysis did not identify any groups with significant differences (Figure \ref{fig:quantitative-sickness}).
Regarding presence, S3 and S4 exhibited higher presence scores than others.
However, there were no significant differences among the scenarios (F(3, 141) = 1.497, $p = .218$, Figure \ref{fig:quantitative-presence}).
In terms of immersion, the ANOVA confirmed significant differences across scenarios (F(3, 141) = 5.22, $p < .01$).
Post-hoc analysis revealed that S3 and S4 exhibited significantly higher immersion compared to S1 (S3: $p < .01$, S4: $p < .05$, Figure \ref{fig:quantitative-immersion}).
Regarding the sense of co-presence, significant differences were observed across scenarios (F(3, 141) = 4.2, $p < .01$).
Specifically, S4 demonstrated a significantly higher sense of co-presence compared to S1 ($p < .05$, Figure \ref{fig:quantitative-copresence}).

In summary, we found that the CIT system, which utilizes chat data from live-streaming concerts to introduce movement reactions, cheering, and singalong sounds in VR concerts, provides statistically higher levels of immersion and co-presence compared to the previous method \cite{Baseline}.

\section{Limitations \& Future Work}
In this paper, we introduced the CIT system, which utilizes chat messages produced by live-streaming concert audiences to synthesize movement, cheering, and singalong reactions in VR concerts.
Through a user study, we found that the CIT system enhances immersion and co-presence.
However, this study has several areas for improvement.
Firstly, while we quantitatively analyzed the viewing experience provided by the CIT, we did not collect participants’ qualitative opinions on individual reactions.
Future research should analyze participants’ feedback on each reaction to gain deeper insights into the user experience enabled by the CIT system.
Secondly, we utilized videos and chat data from live-streaming concerts of the Korean idol group IZ*ONE, and the participants consisted of Koreans in their 20s.
Further research is needed to explore concerts and audiences from a broader range of cultural backgrounds and age groups, enabling a more comprehensive evaluation of the system's effectiveness.
Third, this work does not validate whether the movement and sound reactions in the CIT system accurately reflect audience responses in offline concerts.
In future work, we plan to compare audience reactions in our system with those of offline audiences to ensure their alignment and validity.
Lastly, all the reactions used in this study were created manually.
Automating the creation process of reactions using generative AI could produce more diverse and dynamic audience reactions.
Exploring this possibility represents an intriguing direction for future research.
\section{Conclusion}
Inspired by the observation that some live online concerts utilize multiple delivery methods simultaneously, we proposed a concept to enhance audience experiences in VR concerts by leveraging interaction data from live-streaming concerts.
To determine the audience reactions presented in VR, we first analyzed offline concert footage to identify reactions associated with emotional states.
We then designed the CIT system, which translates chat interactions from large live-streaming audiences into collective artificial reactions for VR concerts.
A user study with the proof-of-concept prototype demonstrated that providing movement and sound reactions through the CIT system significantly enhances immersion and co-presence compared to the previous method.
We hope this study marks an initial step toward enriching user experiences by integrating audience reactions across platforms and fostering interaction in computer-mediated concerts with diverse modalities.

\begin{acks}
This research was supported by Culture, Sports and Tourism R\&D Program through the Korea Creative Content Agency grant funded by the Ministry of Culture, Sports and Tourism in 2025 (Project Name: Development of UX service technology based on new technology convergence content for enjoyment of cultural content by passengers in mobility, Project Number: RS-2024-00441262, Contribution Rate: 80\%).
This research was also partly supported by the Graduate School of Metaverse Convergence support program (IITP-2025-RS-2024-00430997, Contribution Rate: 10\%) and Innovative Human Resource Development for Local Intellectualization program (IITP-2025-RS-2022-00156360, Contribution Rate: 10\%) through the Institute of Information \& Communications Technology Planning \& Evaluation(IITP) grant funded by the Korea government(MSIT).
\end{acks}

\bibliographystyle{ACM-Reference-Format}
\bibliography{sample-base}

\end{CJK}
\end{document}